\documentstyle[12pt]{article}
\begin{document}

\begin{center}
\centering{{\bf INVERSE AC JOSEPHSON EFFECT FOR A FLUXON\\
IN A LONG MODULATED JUNCTION}}
\end{center}

\vspace*{1.0cm}
\begin{center}
\centering{Giovanni Filatrella\\ {\it Physikalisches Institut,
Universit\"{a}t T\"{u}bingen\\D-72076 T\"{u}bingen, Germany\\}{\rm
filatrella@pit.physik.uni-tuebingen.de}}
\end{center}

\smallskip

\begin{center}
\centering{Boris A. Malomed\\ {\it Department of Applied Mathematics,
School of Mathematical Sciences\\Raymond and Beverly Sackler Faculty of
Exact Sciences\\Tel Aviv University\\Ramat Aviv 69978, Israel\\}{\rm
malomed@math.tau.ac.il}}
\end{center}

\smallskip

\begin{center}
\centering{Robert D. Parmentier\\ {\it Dipartimento di Fisica,
Universit\`{a} di Salerno\\I-84081 Baronissi (SA), Italy\\}{\rm
parment@salerno.infn.it}}
\end{center}

\newpage
\begin{center}{\bf ABSTRACT}\end{center}
\vspace*{2truecm}

We analyze motion of a fluxon in a weakly damped ac-driven long Josephson
junction with a periodically modulated maximum Josephson current density.
We demonstrate both analytically and numerically that a pure {\it ac} bias
current can drive the fluxon at a {\it resonant} mean velocity determined
by the driving frequency and the spatial period of the modulation, provided
that the drive amplitude exceeds a certain threshold value. In the range of
strongly ``relativistic'' mean velocities, the agreement between results of
a numerical solution of the effective (ODE) fluxon equation of motion and
analytical results obtained by means of the harmonic-balance analysis is
fairly good; morever, a preliminary PDE result tends to confirm the
validity of the collective-coordinate (PDE-ODE) reduction.  At
nonrelativistic mean velocities, the basin of attraction, in
position-velocity space, for phase-locked solutions becomes progressively
smaller as the mean velocity is decreased.

\newpage
\section{INTRODUCTION}

The `inverse ac Josephson effect' is the generation of dc voltage across
the Josephson junction by ac bias current flowing through it. This effect
was studied in detail for a point-like junction \cite{inverse_effect,FMP},
as well as for shuttle motion of a single fluxon (Josephson vortex) in a
long but finite junction with reflecting edges \cite{shuttle}. In the
present work, we consider a different dynamical regime which may also be
interpreted as the inverse ac Josephson effect: steady progressive motion
of a fluxon in the ac-driven weakly damped long junction with periodically
modulated local parameters. In an experiment, this can be realized by
periodically changing the junction's barrier thickness \cite{Ustinov}; in
the simplest theoretical model it is usually assumed that the only local
parameter which is modulated is the maximum Josephson current, while all
others are kept constant \cite{Mkr}. Thus, the ac-driven periodically
modulated junction is described by the normalized, modified sine-Gordon
equation in the standard notation:
\begin{equation} 
\phi_{tt} - \phi_{xx} + \sin\phi = \epsilon\sin (kx) \sin\phi -\alpha
\phi_{t} + \gamma\sin (\omega t)\, .
\end{equation}
where $\alpha$ is the dissipative constant, $\epsilon$ and $k$ are the
amplitude and wave number of the modulation, and $\gamma$ and $\omega$ are
the amplitude and frequency of the ac bias current driving the junction. We
assume that the ac bias current is distributed uniformly along the
junction, which is a reasonably realistic assumption for overlap- or
annular-geometry junctions \cite{RDP}.

In this work, our objective is to analyze the motion of a fluxon in the
model of Eq.~(1). In the absence of the perturbing factors ($\epsilon
=\alpha =\gamma =0$), the fluxon is described by the well-known solution of
the sine-Gordon equation:
\begin{equation} 
\phi (x,t) = 4\,\tan^{-1}\left[\exp\left(- \frac{x-\xi (t)}{\sqrt{1-
V^{2}}}\right)\right]\, ,
\end{equation}
where $V$ is the fluxon's velocity, and \, $\xi=Vt$ \, is the coordinate of
its center. Following the perturbative approach which has proved to be very
useful in the theory of long Josephson junctions \cite{review}, we will
treat all the terms on the right-hand side of Eq.~(1) as small
perturbations. Then, using energy balance or, equivalently (in this case),
momentum balance, it is straightforward to derive, at the lowest order of
the perturbation theory, the following equation of motion for the fluxon:
\begin{equation} 
\frac{d}{dt}\left(\frac{\dot{\xi}}{\sqrt{1-\dot{\xi}^{2}}}\right) = -
\frac{\epsilon k}{2} \sqrt{1-\dot{\xi}^{2}} \cos (k\xi )-
\frac{\alpha\dot{\xi}}{\sqrt{1-\dot{\xi}^{2}}}+\frac{\pi \gamma}{4} \sin
(\omega t)\, ,
\end{equation}
where, in order to simplify the equation, it is assumed that
\begin{equation} 
\sqrt{1-\dot{\xi}^{2}} \ll \frac{2 \pi}{k} \, ,
\end{equation}
{\em i.e.}, a characteristic width of the fluxon is small in comparison
with the modulation wavelength (in the opposite limit the coefficient in
front of the first term on the right-hand side of Eq.~(3) would be
exponentially small).

Eq.~(3) is an effective equation of motion for a relativistic particle in
the presence of potential, friction, and driving forces, corresponding,
respectively, to the three terms on the right-hand side. Since the driving
force has no dc (constant) part, it may seem that the particle will only
oscillate with zero mean velocity. However, in this work we will
demonstrate that, in the presence of the spatially periodic potential, the
ac drive may support mean progressive motion of the particle, compensating
the dissipation-induced braking. Earlier, it has been predicted
analytically for models of the Toda-lattice (TL) \cite{me} and of the
Frenkel-Kontorova (FK) \cite{Luis} type, and shown numerically for the TL
model in the so-called dual formulation \cite{Jarmo} that ac drive may
support progressive motion of a soliton in weakly damped nonlinear
dynamical lattices; a crucial circumstance which made the effect possible
was the spatial periodicity of the lattice. Moreover, the stable
propagation of ac-driven solitons in a model electric transmission line
described by the damped dual TL equation was detected experimentally in
the work \cite{Tom}. Note that for the TL in its usual form the effect does
not take place because the ac drive can provide only for the balance of
energy, but not for that of the momentum, while in the dual formulation the
soliton's momentum is identically zero, and energy balance becomes
sufficient \cite{Cai}. In the FK model the balance of energy alone is, in
principle, sufficient since the momentum is not conserved in this system
even in the absence of dissipation and drive, but the threshold predicted
by the theory \cite{Luis} for the effect proves to be so high that it
would be hard to observe in a numerical experiment \cite{Cai}.

In the present work, we are dealing with the continuum model of Eq.~(1).
However, in the presence of a periodic potential, the ac drive allows the
soliton in this damped continuum system to move at a nonzero mean velocity,
the underlying mechanism being essentially the same as in the discrete
lattices. Indeed, if the soliton is moving with a mean velocity
$\overline{V}$ through the periodic potential, this gives rise to the
temporal frequency $k\overline{V}$ (recall that $k$ is the wave number
which determines the spatial periodicity). The condition
\begin{equation} 
k\overline{V} = \omega \, ,
\end{equation}
where $\omega$ is the driving frequency, may give rise to a resonant
transfer of energy from the ac driving field to the soliton. In turn, this
supply of energy gives the soliton a chance to compensate the dissipative
losses. In the case of the long Josephson junction, the mean dc voltage
across the junction is proportional to the mean velocity of the fluxon.
Thus, supporting the fluxon's progressive motion by the ac bias current may
be regarded as another manifestation of the inverse ac Josephson effect.

In section~2 of this work, we will develop a fully analytical approach to
investigate the ac-driven motion of the fluxon in the ``ultrarelativistic''
case, {\em i.e.}, when the mean velocity is close enough to the junction's
limit (Swihart) velocity (which is equal to one in the notation adopted).
In section~3, we will display results of the full numerical solution of the
ODE, Eq.~(3).

\section{ANALYTICAL CONSIDERATIONS}

An analytical approach to Eq.~(3) is possible if one assumes that the small
parameters $\alpha$ and $\gamma$ are ``very small'', while $\epsilon$ is
``not so small'' (the physically relevant range for $\epsilon$ is
$0~\leq~\epsilon~\leq~1$). In other words, we will consider, in the lowest
approximation, motion of the free relativistic particle in the periodic
potential, and then the friction and driving forces will be taken into
account as additional small perturbations. Another fundamental assumption
which makes the analytical consideration possible is that the particle is
moving at ultrarelativistic velocities, {\em i.e.}, we will set
\begin{equation} 
\dot{\xi} \equiv 1-w^{2} \,\, , w^{2} \ll 1 \, .
\end{equation}
Insertion of Eq.~(6) into Eq.~(3) leads, after simple algebra, to the
following reduced equation of motion for the variable $w$:
\begin{equation} 
\frac{dw}{dt} = \epsilon kw^{3}\cos (k\xi)+\alpha w -\frac{\pi\gamma}
{2\sqrt{2}}\, w^{2} \sin (\omega t)\, .
\end{equation}

In what follows below, we will concentrate on the case when the
``particle'' moves with a nearly constant velocity. In this case, the
variable \,$\xi$\, in the argument of the cosine in the first term on the
right-hand side of Eq.~(7) may be replaced, in the lowest approximation, by
\begin{equation} 
\xi^{(0)} \equiv \overline{V} t+\xi_{0} \, ,
\end{equation}
where $\overline{V}$ is the mean velocity (the same as in Eq.~(5)), and
\,$\xi_{0}$\, is a constant phase shift between the particle's mean law of
motion and the time-periodic driving force. Finally, introducing the
renormalized time $\tau \equiv k\overline{V} t$, we obtain the lowest-order
equation of motion in the form (with $\delta \equiv k\xi_{0}$)
\begin{equation} 
\frac{dw}{d\tau} = \frac{\epsilon}{\overline{V}}\, w^{3} \cos (\tau+\delta)
+ \frac{\alpha}{k \overline{V}}\, w - \frac{\pi \gamma}{2\sqrt{2}k}\, w^{2}
\sin (\frac{\omega }{k\overline{V}}\,\tau)\, .
\end{equation}

Now, we will make use of the fundamental assumption mentioned above, {\em
viz.}, that the driving and friction terms may be regarded as small in
comparison with the first (potential) term on the right-hand side of Eq.~
(9). In the zero'th-order approximation, we drop the small terms, obtaining
from Eq.~(9)
\begin{equation} 
\frac{dw}{d\tau} = \frac{\epsilon}{\overline{V}}\, w^{3} \cos (\tau+\delta)
\, .
\end{equation}
The exact solution of Eq.~(10) is
\begin{equation} 
w = \sqrt{\frac{\Delta}{1-\frac{2 \epsilon\Delta}{\overline{V}}\sin (\tau +
\delta)}} \,\, ,
\end{equation}
where $\Delta$ is an arbitrary constant. According to Eq.~(6), $w^{2}~\ll~
1$, which implies that $\Delta~\ll~1$. Moreover, the physically significant
range for $\epsilon$ is $\epsilon \leq 1$; and finally,
$\overline{V}~\approx~1$. Consequently, the coefficient of the sine term in
the denominator of Eq.~(11) is small enough to replace this expression by
its expansion,
\begin{equation} 
w = \sqrt{\Delta} \left[ 1+\frac{\epsilon \Delta}{\overline{V}} \sin (\tau
+\delta) \right] \, .
\end{equation}

Eqs. (6) and (12) describe the motion of a particle at the mean velocity
$\overline{V} = 1-\Delta$, with superimposed small oscillations of the
velocity corresponding to the second term in the expression (12). Inserting
this \,$\overline{V}$\, into the resonance condition of Eq.~(5), we find
the following relation between the driving frequency $\omega$ and the
corresponding {\it resonant} value of the parameter $\Delta$:
\begin{equation} 
\omega = k(1-\Delta)\, .
\end{equation}
Obviously, in a real physical situation the system chooses itself the
resonant value of $\Delta$ corresponding to given $\omega$. However, in the
analytical consideration that follows below it will be more convenient to
regard $\Delta$ as a given arbitrary parameter (small enough), and then to
match the frequency to it according to Eq. (13).

A resonant drive can support progressive motion of a particle in a lossy
periodic medium if the drive amplitude exceeds a certain minimum ({\it
threshold}) value proportional to the friction coefficient \cite{me}. In
fact, the threshold value of the amplitude, regarded as a function of the
driving frequency, is the basic characteristic of such ac-driven motion of
the soliton in the damped system \cite{me,Jarmo}. In the present case, the
amplitude of the driving force is the parameter $\gamma$. A convenient,
approximate technique for calculating its threshold value, $\gamma_{{\rm
thr}}$, is to use the idea of harmonic balance \cite{Minorsky}, {\em i.e.},
we insert the approximate solution given by Eq. (12) into the last two
terms on the right-hand-side of Eq. (9), adjusting the coefficients in
these terms so that the zero'th harmonic (and, in general, as many
harmonics as possible) vanishes. The result of such a calculation (for the
zero'th harmonic) is
\begin{equation} 
\gamma=\frac{4\sqrt{2}\alpha}{\pi\epsilon}\Delta^{-3/2}/|\cos(\delta)|\,.
\end{equation}
Recalling that\, $\delta$\, is an arbitarary phase shift, it is apparent
that the threshold corresponds to $|\cos(\delta)|=1$ in Eq. (14), {\em
i.e.},
\begin{equation} 
\gamma_{{\rm thr}} = \frac{4\sqrt{2}\alpha}{\pi\epsilon}\Delta^{-3/2} \, .
\end{equation}

This is our main analytical result. The proportionality of $\gamma_{{\rm
thr}}$ to the dissipation constant $\alpha$, as well as the inverse
proportionality to $\epsilon$, are obvious. A nonobvious feature of
Eq.~(15) is the power\, $-\frac{3}{2}$\, of $\Delta$ (it is evident, of
course, that this power must be negative). In the next section, comparing
with results of direct numerical simulations of the equation of motion,
Eq.~(3), will show that this particular value of the power, as well as the
numerical prefactor in Eq.~(15), are quite reasonably accurate.

\section{NUMERICAL RESULTS}

In the numerical integration of Eq.~(3), we fixed the driving frequency to
$\omega = 1$. The justification for this step is provided by noting that
$\omega$ can be scaled out of Eq.~(3), {\em i.e.}, set to unity, by
defining the following re-scaled variables: $t' \equiv \omega t,\, \xi'
\equiv \omega \xi$ (which leaves $\dot{\xi}$ unchanged), $k' \equiv
k/\omega,\, \alpha' \equiv \alpha /\omega,\, \gamma' \equiv \gamma
/\omega$, and $\epsilon' \equiv \epsilon$. Such a re-scaling is valid
provided that condition~(4) continues to hold, but it obviously neglects
any possible frequency-dependent resonance effects with the background
oscillation (for the case when the driving frequency is close
to the junction's plasma frequency, the resonance effects were analyzed
in detail in \cite{Borya}). Then, with $\Delta$ taken as an arbitrary
parameter (small enough), we calculated the modulation wave number
\,$k$\, from the resonant relation~(13). The mode of simulations chosen
was as follows: with all the parameters but the drive amplitude
\,$\gamma$\, fixed, we reduced the value of \,$\gamma$\, quasi-
adiabatically until the solution dropped out of synchronism with the
driver; the smallest value of $\gamma$ so obtained was taken as
\,$\gamma_{{\rm thr}}$\,. The numerical results obtained are summarized
in Figs. 1 and 2.

Figs. 1(a) and 1(b) show, respectively, the inverse and direct
proportionality of $\gamma_{{\rm thr}}$ to the parameters $\epsilon$ and
$\alpha$. The analytical prediction of Eq.~(15) is qualitatively well
confirmed, even though there is a small quantitative discrepancy between
analytical and numerical results. Fig.~2, instead, shows the numerical
dependence on $\Delta$. Here, we see that, for small enough values of
$\Delta$, the agreement between analytical and numerical results is quite
good. The agreement gradually worsens as $\Delta$ increases, which is not
surprising in view of the fact that $\Delta \ll 1$ is a fundamental
assumption of the ultrarelativistic analysis leading to Eq.~(15).

In the extreme nonrelativistic limit, {\em i.e.}, $\Delta \rightarrow 1$,
which implies $\overline{V} \rightarrow 0$, Eq.~(3) becomes, with trivial
modifications, the equation for the damped, ac-driven, simple pendulum. The
existence of a threshold value of the drive amplitude for phase-locked
motion in this case was recently derived by Filatrella {\em et al.}
\cite{FMP} using an energy-balance argument. The solid curve in Fig.~2
shows the drive-amplitude threshold calculated for this case; as is evident
from the figure, the numerical solution of Eq.~(3) increasingly approaches
this curve as $\Delta \rightarrow 1$.

In Ref. \cite{FMP}, the numerical solutions of the ``nonrelativistic''
equation showed a gradual shrinkage, and eventual disappearance, with
decreasing $\omega$, of the basin of attraction in position-velocity
space of the characteristic phase-locked motion. We have seen the same
phenomenon, which, from Eq.~(13), corresponds to the situation $\Delta
\rightarrow 1$, in our present numerical solutions of Eq.~(3). We suggest
that this phenomenon might be attributable to the co-existence, in this
region of parameter space, of chaotic trajectories in the underlying
dynamics.

A crucial step in our analysis has been the reduction of the system
description from the PDE of Eq.~(1) to the ODE of Eq.~(3). In order to
check that the existence of ac-driven phase-locked states is not just an
artifact of this reduction we have performed a very preliminary numerical
study of Eq.~(1), in which we have seen clear evidence, for several
different parameter combinations, that the phenomenon is also present in
the PDE \cite{Giofil}. In these preliminary tests we applied periodic
boundary conditions to Eq.~(1), with the junction length chosen to
contain one period of the spatial modulation, {\em i.e.}, $L = 2 \pi/k$,
with $k$ given by Eq.~(13). The PDE was discretized in space using a
three-point approximation for the second spatial derivative, and
integrated in time using a fourth-order Runge-Kutta routine. Eq.~(2),
with $\xi = 0$ and $V = 0$, was used as the initial condition. In one
such run, with $\alpha = 0.005, \, \epsilon = 1.0, \, \gamma = 0.53, \,
\omega = 1.0, \, \Delta = 0.1$ and with a time step of 0.02 and a spatial
grid of 100 points, the fluxon continued to propagate for at least 2500
normalized time units. Further detailed support for the very broad range
of validity of the collective-coordinate (PDE-ODE) reduction in the
periodically modulated sine-Gordon equation (without loss or driving
terms) has recently been presented by S\'{a}nchez {\em et al.} \cite{Angel}.

\section{CONCLUSIONS}

In this work, we have demonstrated that a soliton in a periodically
inhomogeneous continuum lossy medium can be driven at a nonzero mean
velocity by an ac driving force. We have also found a good agreement
between the analytical approximation based on harmonic balance and direct
numerical simulations of the effective (ODE) equation of motion for the
soliton. A noteworthy result is that the soliton can be easily driven in
the ``ultrarelativistic'' region, but less easily at moderately
relativistic and nonrelativistic values of the mean velocity. In this
connection, one might reasonably ask, if the basin of attraction for
ac-driven phase-locked states is small, whether or not such states can be
of any potential physical interest, {\em e.g.}, for practical applications.
Our answer to this question is positive: the fact that an experimentalist
might have to ``prepare'' his system, {\em e.g.}, by initially applying a
combination dc + ac drive, and then gradually reducing the dc bias to zero,
is not a serious impediment. What is important is that the basin of
attraction be sufficiently large so that, once having arrived in an
ac-driven phase-locked state (no matter how), the system be sufficiently
stable against thermal noise and other perturbations, for a sufficiently
long time to make measurements. The technique of cell-to-cell mapping
\cite{Hsu,NFP}, applied to Eq.~(3), might provide a useful tool for
obtaining some initial insight into this problem.

Finally, it is relevant to mention that in this work we have confined
ourselves to consideration of the simplest resonance only. Higher-order
resonances may also take place, in which $mk\overline{V}=n\omega$, with
incommensurable integers $m$ and $n$, {\em cf}. Eq. (5). It is natural to
expect that the threshold amplitudes for the higher resonances will be
considerably larger than for the fundamental resonance considered here.

\section*{Acknowledgements}

We are grateful to David Cai and Angel S\'{a}nchez for stimulating
discussions. B.A.M. thanks the Physics Department of the University of
Salerno for hospitality during the visit that originated this work.
Financial support from the EU under contract no. SC1-CT91-0760 (TSTS) of
the ``Science'' program and contract no. ERBHBGCT920215 of the ``Human
Capital and Mobility'' program, from MURST (Italy), and from the Progetto
Finalizzato ``Tecnologie Superconduttive e Criogeniche'' del CNR (Italy) is
gratefully acknowledged.

\newpage
\section*{Figure Captions}

\noindent
Fig. 1. Dependence of the threshold value of the driving force,
$\gamma_{{\rm thr}}$, on the parameters: (a) $\epsilon$ (modulation
coefficient), with $\alpha = 0.001,\, \omega = 1.0,\, \Delta = 0.1$; (b)
$\alpha$ (dissipation coefficient), with $\epsilon = 1.0,\, \omega = 1.0,\,
\Delta = 0.1$. Dash-dotted lines: prediction of Eq.~(15); asterisks:
numerical solutions of Eq.~(3). Solid lines connecting the asterisks are a
guide to the eye only.

\noindent
Fig. 2. Dependence of the threshold value of the driving force,
$\gamma_{{\rm thr}}$, on the velocity parameter, $\Delta$, with $\alpha =
0.001,\, \omega = 1.0,\, \epsilon = 0.5$. Dash-dotted line: prediction of
Eq.~(15); asterisks: numerical solution of Eq.~(3); solid curve:
analytical prediction of a nonrelativistic theory (see text). Solid lines
connecting the asterisks are a guide to the eye only.

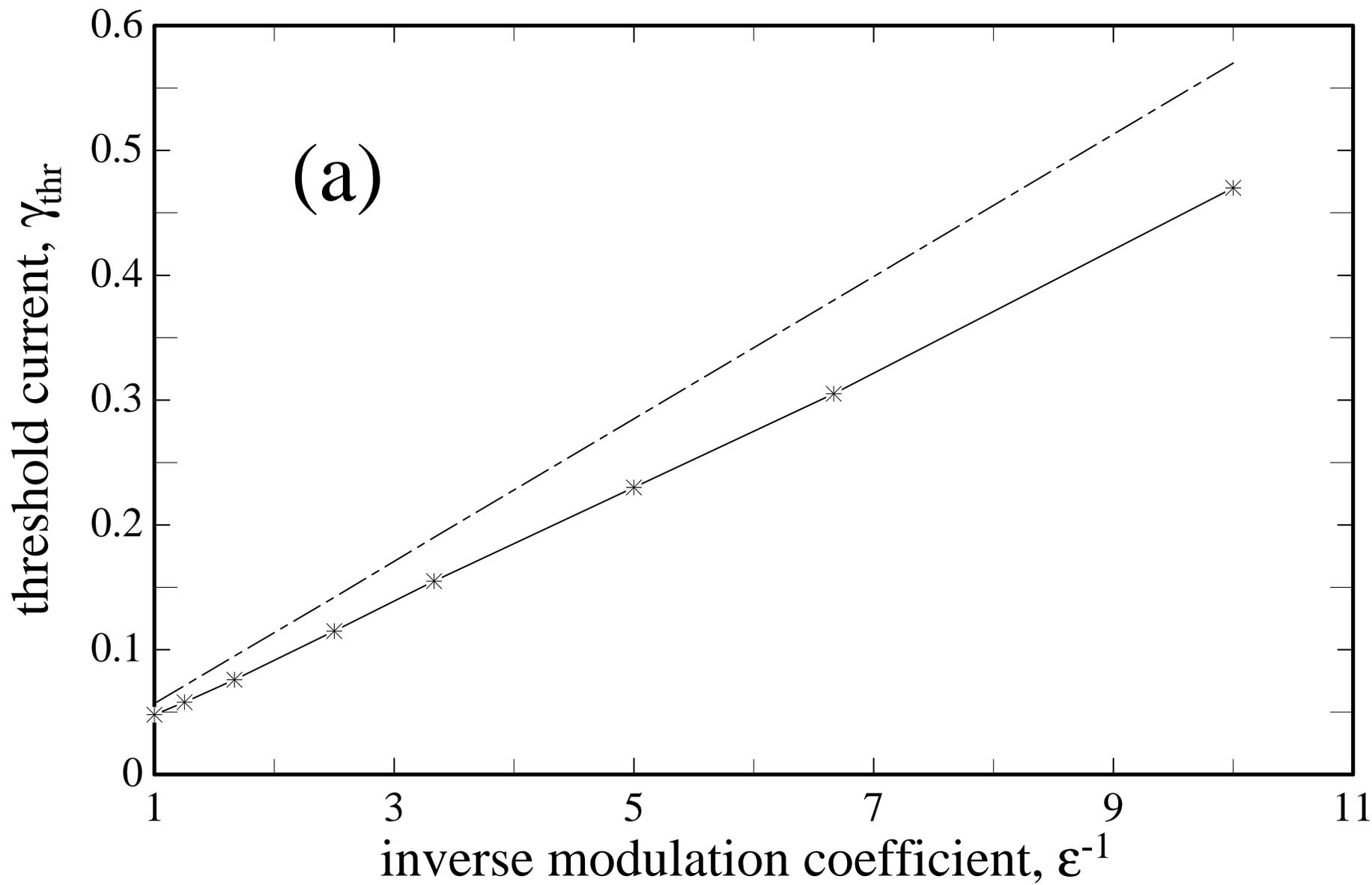
\end{document}